\documentclass[12pt,a4paper,reqno]{amsart}
\usepackage{xypic,a4wide,amsmath,amssymb,amsthm}
\usepackage[utf8x]{inputenc}

\usepackage[T1]{fontenc}

\usepackage{amssymb}
\usepackage{amsmath}

\usepackage[dvips]{graphicx}
\usepackage{array}

\title{Lax pair representation and Darboux transformation of NC Painlev\'e-II equation }

\author{M. Irfan}
\address{
 (1)-Department of Mathematics, Universit\'e d'Angers,\\
2 Boulevard Lavoisier, 49045 Angers Cedex 01, France\\
(2)-Institute of Geology, University of the Punjab, Pakistan} \email{mahirfan@yahoo.com} \urladdr{}

\begin{document}

\maketitle

\begin{abstract}
The extension   of Painlev\'e equations to noncommutative spaces has been considering extensively  in the theory of integrable systems and it is also interesting to explore some remarkable aspects of these equations such as Painlev\'e property, Lax representation, Darboux transformation  and their connection to well know integrable equations. This paper is devoted to the Lax formulation, Darboux transformation and  Quasideterminant solution of noncommutative Painlev\'e second equation which is recently introduced by V. Retakh and V. Rubtsov.  

\end{abstract}

\section{Introduction}
The Painlev\'e equations were discovered by Painlev\'e and his colleagues when they were  classifying  the  nonlinear second-order ordinary differential equations with respect to their solutions \cite{r1}.The importance of Painlev\'e equations from  mathematical point of view is because of their frequent appearance in the  various areas of physical sciences including plasma physics, fiber optics, quantum gravity and field theory, statistical mechanics, general relativity and nonlinear optics. The classical Painlev\'e equations are regarded as completely integrable equations and obeyed the Painlev\'e test \cite{r5,r11,r12}.
These equations are subjected to the some properties such as linear representation, hierarchy, Darboux transformation(DT)  and Hamiltonian structure  because of their reduction from integrable systems, i.e, Painlev\'e second (P-II) equation arises as reduction of KdV equation  \cite{r4,MA} .\\
The Noncommutative(NC) extension of Painlev\'e equations  is  quite interesting  in order to explore their properties  which they  possess on ordinary spaces.
NC spaces are characterized by the noncommutativity of the spatial co-ordinates, if $ x^{\mu} $ are the space co-ordinates then the noncommutativity is defined by
 $ [x^{\mu},x^{\nu}]_{\star}=i\theta^{\mu} $
where parameter $ \theta^{\mu\nu}$ is anti-symmetric tensor and Lorentz invariant and $ [x^{\mu},x^{\nu}]_{\star}$ is commutator under the star product. NC field theories on flat spaces are given by the replacement of ordinary
products with the Moyal-products and realized as
deformed theories from the commutative ones. Moyal product for ordinary fields $ f(x)$  and $ g (x)$ is  explicitly defined by
\[f(x) \star g(x)=exp(\frac{i}{2}\theta^{\mu\nu} \frac{\partial}{\partial x^{'\mu}}\frac{\partial}{\partial x^{''\nu}})f(x^{'})g(x^{''})_{x=x^{'}=x^{''}}\]
\[=f(x)g(x)+\frac{i}{2}\theta^{\mu\nu}\frac{\partial f}{\partial x^{'\mu}}\frac{\partial g}{\partial x^{''\nu}}+ \mathcal{O}(\theta^2).\]this product obeys associative property
$f\star ( g \star h)= (f \star g) \star h$,
if we apply the commutative limit
$\theta^{\mu \nu} \rightarrow 0 $
then above expression will reduce to ordinary product as   $ f \star g = f. g.$ \\
We are familiar that Lax equation is a nice representation of integrable systems, the form of  Lax equation on  deformed space is  the same  as it has on ordinary space, here  ordinary product is replaced by the star product. The Lax equation involves two linear operators, these operators may be differential operators or matrices \cite{r13}-\cite{r24}.
If $A$ and $B$ are the linear operators then Lax equation is given by  $ A_{t} = [B,A]_{\star}$  where $[B,A]_{\star}$
is commutator under the star product, this Lax pair formulasim is also helpful to construct the DT of integrable systems. Now consider a linear system $\Psi_{x}=A(x,t)\Psi$ and $\Psi_{t}=B(x,t)\Psi $, the compatibility of this system yields $ A_{t}-B_{x}= [B,A]_{\star}$  which is called zero curvature condition \cite{e2} -\cite{4}, further let we express the commutator $[,]_{-}$ and anti-commutator $[,]_{+}$ without writing the $ \star $ as subscript then zero curvature condition may be written as $ A_{t}-B_{x}= [B,A]_{-}$. \\

In this paper, i have applied  Lax pair formalism for the representation of NC P-II equation and this work also involves the explicit description of DT of this equation. Finaly, i  derive the multi-soliton solution of NC P-II equation in terms of quasideterminants.\\
 
\subsection{Linear representation} 
Many integrable systems possess the  linear representation on ordinary as well as on NC spaces, this representation is also known by the Lax representation, matrix P-II equation on ordinary space has this kind of representation  \cite{6}. Here, we will see that how the NC P-II equation arises from the  compatibility condition of following linear systems
\[ \Psi_{\lambda}=A(z;\lambda)\Psi , \Psi_{z}=B(z;\lambda)\Psi ,\]
where $ A(z;\lambda) $ and $ B(z;\lambda) $ are matrices. The compatibility condition
$ \Psi_{z \lambda} = \Psi _{\lambda z}$
implies
\begin{equation}\label{9}
 A_{z}-B_{\lambda}= [B,A]_{-}
\end{equation}
above expression is similar to zero curvature condition and an alternative linear representation of NC P-II equation with matrices $ A $ and $ B $ defined as under

\[ A = \begin{pmatrix}

8  i\lambda^{2} +  i v^{2}-2i z & -i v_{z}+\frac{1}{4}C \lambda^{-1}-4\lambda v  \\
i v_{z}+\frac{1}{4}C \lambda^{-1}-4\lambda v  &  -8  i\lambda^{2} - i v^{2}+2i z
 \end{pmatrix}\]
\[ B =  \begin{pmatrix}

-2i \lambda  & v \\
v & 2i \lambda
 \end{pmatrix}\]
where the $ \lambda $ is a commuting parameter and $C$ is a constant. Now we can easily evaluate the following values
\[A_{z}-B_{\lambda}= \begin{pmatrix}
  i v_{z}v+iv v_{z} & -i v_{zz}-4\lambda v_{z}  \\
i v_{zz}-4\lambda v_{z}  &  -i v_{z}v-iv v_{z}
\end{pmatrix}\]
and
\[BA-AB=\begin{pmatrix}
a & -b-4\lambda v_{z} \\
b -4\lambda v_{z} &  -a
\end{pmatrix}\]
where\[a=i v_{z}v+ iv v_{z}\] \[b=2iv^{3}-2i[z,v]_{+} +iC \]
now by using the above values  in (\ref{9}) we have the following expression
\[\begin{pmatrix}
0 & -i v_{zz}+2iv^{3}-2i[z,v]_{+} +iC \\
i v_{zz}-2iv^{3}+2i[z,v]_{+} -iC & 0
\end{pmatrix}=0\] and finally we get
\begin{equation}\label{p2}
 v_{zz}=2v^{3}-2[z,v]_{+} +C
\end{equation}
 equation  (\ref{p2}) is similar to the NC P-II equation which is introduced by V. Retakh amd V. Rubtsov \cite{7}.
\section{NC Symmetric Functions and Lax Representation } This section consists the Lax representation of the set of three equations of functions  $ u_{0} $, $ u_{1} $ and $ u_{1}$ , this set is equivalent to the NC P-II equation \cite{7} .
Now consider the linear system \[L_{t}\psi=\lambda \psi \]  the time evolution of $\psi$ is given by \[ \psi_{t}= P\psi \] and the above system is equivalent to the Lax equation 
\begin{equation}\label{L2}
 L_{t}=[P,L]_{-}
\end{equation}
here $\lambda$ is a spectral parameter and $\lambda_{t}=0$. Now we take the Lax pair $ L,P $ in the following form
\[L= \left(
\begin{array}{c|c|c}

 L_{1} & O & O \\ \hline
 O & L_{2} & O \\ \hline
 O & O & L_{3}
\end{array}
\right)
\]

and 
\[P= \left(
\begin{array}{c|c|c}

 P_{1} & O & O \\ \hline
 O & P_{2} & O \\ \hline
 O & O & P_{3}
\end{array}
\right)
\]
where 
\[  L_{1} = \begin{pmatrix}
1 & 0 \\
-v_{0} & -1
\end{pmatrix},   L_{2} = \begin{pmatrix}
1 & 0 \\
-v_{1} & -1
\end{pmatrix},   L_{3} = \begin{pmatrix}
-1 & 0 \\
-v_{2} & 1
\end{pmatrix} \]
and the elements of matrix $ P$ are given by \[  P_{1} = \begin{pmatrix}
\rho_{1} & 0 \\
0 & -\rho_{1}
\end{pmatrix},   L_{2} = \begin{pmatrix}
-\rho_{2} & 0 \\
0 & \rho_{2}
\end{pmatrix},   L_{3} = \begin{pmatrix}
-1 & 0 \\
\frac{1}{2}\sigma & 1
\end{pmatrix} \]
 where 
\[\rho_{1}=-v_{2}-\frac{1}{2}\alpha_{0}v^{-1}_{0} , \rho_{2}=-v_{2}+\frac{1}{2}\alpha_{1}v^{-1}_{1} ,\sigma=v_{0}-v_{1}+2v_{2}\]
and 
\[O=  \begin{pmatrix}
0 & 0 \\
0 & 0
\end{pmatrix}.\]
When the above Lax matrices $ L$ and $ P$ are subjected to the Lax equation (\ref{L2})
 we get
\[v^{'}_{0}=v_{2}v_{0}+v_{0}v_{2} + \alpha_{0}\]
\[v^{'}_{1}=-v_{2}v_{1}-v_{1}v_{2} + \alpha_{1}\]
\[v^{'}_{2}=v_{1}-v_{0} \] the above system can be reduced to NC P-II equation (\ref{p2}) by eliminating $v_{0} $ and $v_{1} $. For this Lax representation of symmetric functions , the quasideterminants  $|I+\mu L|_{ii}$  can not be expressed in terms of the expansion of symmetrics functions \cite{NCSYM}.
\section{A Brief Introduction of Quasideterminants}
This section is devoted to a brief review of quasideterminants introduced by Gelfand
and Retakh  \cite{Gel Ret}. Quasideterminants are the replacement for the determinant for matrices with noncommutative entries and these determinants plays very important role to construct the multi-soliton solutions of NC integrable systems \cite{r25,r26}, by applying the Darboux transformation. Quasideterminants are not just a noncommutative generalization of usual commuta-
tive determinants but rather related to inverse matrices, quasideterminants for the square matrices are defined as
if $A = a_{ij}$ be a $ n \times n $ matrix and $B = b_{ij}$ be the inverse matrix of A. Here all
matrix elements are supposed to belong to a NC ring with an associative
product.
Quasideterminants of $A$ are defined formally as the inverse of the elements of $B = A^{-1}$
\[ |A|_{ij}=b^{-1}_{ij} \] this expression under the limit $ \theta^{\mu \nu} \rightarrow 0 $ , means entries of $A$ are commuting, will reduce to \[ |A|_{ij}= (-1)^{i+j}\frac{detA}{detA^{ij}} \]
where $A^{ij}$ is the matrix obtained from $ A $ by eliminating  the $ i$-th row and the $ j$-th column.
We can write down more explicit form of quasideterminants. In order to see it, let us
recall the following formula for a square matrix

\begin{equation} A =
\begin{pmatrix}
 A & B \\
C & D
\end{pmatrix}^{-1}
=\begin{pmatrix}
A-BD^{-1}C)^{-1} & -A^{-1}B(D-CA^{-1}B)^{-1} \\
-(D-CA^{-1}B)^{-1} CA^{-1} & (D-CA^{-1}B)^{-1}
\end{pmatrix}
\end{equation}
where $A$ and $D$ are square matrices, and all inverses are supposed to exist. We note that
any matrix can be decomposed as a $
2\times 2$ matrix by block decomposition where the diagonal
parts are square matrices, and the above formula can be applied to the decomposed $2 \times 2$
matrix. So the explicit forms of quasideterminants are given iteratively by the following
formula

\[ \vert A \vert _{ij}=a_{ij}-\Sigma_{p\neq i , q\neq j}  a_{iq} \vert {A}^{ij} \vert^{-1} _{pq} a_{pj}  \]
the  number of quasideterminant of a given matrix will be equal to the numbers of its elements for example a matrix of order $3$ has nine quasideterminants. It is sometimes convenient to represent the quasi-determinant as follows
\[  \vert A \vert_{ij}=\begin{vmatrix}
a_{11} &  \cdots  &  a_{1j}          & \cdots  & a_{1n}\\
\vdots & \vdots   & \vdots           & \vdots  & \vdots\\
a_{i1} & \cdots   & \boxed{a_{ij}}    & \cdots  & a_{in}\\
\vdots & \vdots   & \vdots           & \vdots  & \vdots\\
a_{in} & \cdots   & a_{ni}           & \cdots  & a_{nn}
\end{vmatrix}. \]
Let us consider  examples of matrices with order $2$ and $3$, for $2\times2$ matrix

\[ A = \begin{pmatrix}
          a_{11} & a_{12} \\
          a_{21} & a_{22} \\
       \end{pmatrix} \]
now the quasideterminats of this matrix are given below
\[  \vert A \vert _{11}=\begin{vmatrix}
{\boxed{a_{11}}} & a_{12} \\
a_{21} & a_{22}
\end{vmatrix} = a_{11} - a_{12} a^{-1}_{22} a_{21} \]

 \[  \vert A \vert _{12}=\begin{vmatrix}
a_{11} &  {\boxed{a_{12}}}\\
a_{21} & a_{22}
\end{vmatrix} = a_{12} - a_{22} a^{-1}_{21} a_{12} \]

\[  \vert A \vert _{21}=\begin{vmatrix}
a_{11} & a_{12}\\
 {\boxed{a_{21}}} & a_{22}
\end{vmatrix} = a_{21} - a_{11} a^{-1}_{12} a_{22} \]

 \[  \vert A \vert _{22}=\begin{vmatrix}
a_{11} & a_{12}\\
a_{21}& {\boxed{a_{22}}}
\end{vmatrix} = a_{22} - a_{21} a^{-1}_{11} a_{12}. \]
 The number of quasideterminant of a given matrix will be equal to the numbers of its elements for example a matrix of order $3$ has nine quasideterminants.
Now we consider the example of $3\times3$ matrix, its first quasidetermints can be evaluated in the following way\\

$ \vert A \vert _{11}= \begin{vmatrix}
{\boxed{a_{11}}} & a_{12} & a_{13}\\
 a_{21} & a_{22} & a_{23}\\
 a_{31} & a_{32} & a_{33}
\end{vmatrix}= a_{11}-a_{12} M a_{21}-a_{13} M a_{21}-a_{12} M a_{31}-a_{13} M a_{31}$\\
where $ M =\begin{vmatrix}
{\boxed{a_{22}}} &  a_{23}\\
a_{32} & a_{33}
\end{vmatrix}^{-1} ,$ similarly we can evaluate the other eight quasideterminants of this matrix .

\section{Darboux transformation}

Darboux transformations play a very important role to construct the multi-soliton solutions of integrable systems these transformations are deduced from the given linear system of integrable systems. To derive the DT of NC P-II equation we consider its linear systems with the column vector $\psi=
\begin{pmatrix}
\chi  \\
 \Phi
\end{pmatrix}$.
Now the linear system will become
\begin{equation}\label{di}
 \begin{pmatrix}
\chi  \\
 \Phi
\end{pmatrix}_{\lambda} =\begin{pmatrix}
8  i\lambda^{2} +  i v^{2}-2i z & -i v_{z}+\frac{1}{4}C \lambda^{-1}-4\lambda v  \\
i v_{z}+\frac{1}{4}C \lambda^{-1}-4\lambda v &  -8 i\lambda^{2} - i v^{2}+2i z
\end{pmatrix}
\begin{pmatrix}
 \chi  \\
 \Phi
\end{pmatrix}
\end{equation}
\begin{equation}\label{d2}
 \begin{pmatrix}
 \chi  \\
\Phi
\end{pmatrix}_{z} =
\begin{pmatrix}
 -2i \lambda  & v \\
v & 2i \lambda
\end{pmatrix}
\begin{pmatrix}
 \chi  \\
 \Phi
\end{pmatrix}.
\end{equation} The standard transformations \cite{DTI,DT3,DT4} on $\chi$ and $\Phi$ are given below
\begin{equation}\label{d3}
\chi \rightarrow \chi[1]=\gamma \Phi-\gamma_{1}\Phi_{1}(\gamma_{1})\chi_{1}^{-1}(\gamma_{1})\chi
\end{equation}
\begin{equation}\label{d4}
\Phi \rightarrow \Phi[1]=\gamma \chi-\gamma_{1}\chi_{1}(\gamma_{1})\Phi_{1}^{-1}(\gamma_{1})\Phi
\end{equation}
where $\chi$ , $\Phi$ are arbitrary solutions at  $\gamma$ and  $\chi_{1}(\gamma_{1})$ ,  $\Phi_{1}(\gamma_{1})$ are the particular solutions at $\gamma=\gamma_{1}$ of equations (\ref{di}) and (\ref{d2}), these equations will take the following forms under the transformations (\ref{d3}) and (\ref{d4})
\begin{equation}\label{d5}
 \begin{pmatrix}
  \chi[1]  \\
 \Phi[1]
 \end{pmatrix}_{\lambda} =
\begin{pmatrix}
 8i\lambda^{2} + i v^{2}[1]-2i z & -i v_{z}[1]+\frac{1}{4}C \lambda^{-1}-4\lambda v[1]  \\
iv_{z}[1]+\frac{1}{4}C \lambda^{-1}-4\lambda v[1]  &  -8 i\lambda^{2} - i v^{2}[1]+2i z
\end{pmatrix}
\begin{pmatrix}
 \chi[1]  \\
 \Phi[1]
\end{pmatrix}
\end{equation}
\begin{equation}\label{d6}
 \begin{pmatrix}
  \chi[1]  \\
 \Phi [1]
 \end{pmatrix}_{z} =
\begin{pmatrix}
 -2i \lambda  & v[1] \\
v[1] & 2i \lambda
\end{pmatrix}
\begin{pmatrix}
 \chi[1]  \\
 \Phi[1]
\end{pmatrix}.
\end{equation}
Now from  (\ref{d2}) and equation (\ref{d6}) we have the following expressions
\begin{equation}\label{d7}
\chi_{z}=-i\lambda\chi+v\Phi
\end{equation}

\begin{equation}\label{d8}
\Phi_{z}=i\lambda\Phi+v\chi
\end{equation}
and
\begin{equation}\label{d9}
\chi_{z}[1]=-i\lambda\chi[1]+v[1]\Phi[1]
\end{equation}
\begin{equation}\label{d10}
\Phi_{z}[1]=i\lambda\Phi[1]+v[1]\chi[1].
\end{equation}
Now  substituting the transformed values  $\chi[1] $ and  $\Phi[1] $  in equation (\ref{d9}) and then after using the  (\ref{d7}) and (\ref{d8}) in resulting equation, we get
\begin{equation}\label{d11}
v[1]=\Phi_{1}\chi_{1}^{-1}v \Phi_{1}\chi^{-1}_{1}.
\end{equation}
 Equation (\ref{d11}) represents the Darboux transformation of NC P-II equation, where $v[1]$ is a new solution of NC P-II equation, this shows that how the new solution is related to the seed solution $v$. By applying the DT iteratively we can construct the multi-soliton solution of NC P-II equation.
\subsection{Quasideterminant solutions}The transformations (\ref{d3}) and (\ref{d4}) may be expressed in the form of quasideterminants ,
first consider the equation (\ref{d3}) in follows form \[ \chi [1]= \gamma_{0} \Phi_{0} -\gamma_{1} \Phi_{1}(\gamma_{1}) \chi^{-1}_{1}(\gamma_{1})\chi_{0}  \]or
\begin{equation}\label{qd1}
\chi [1]=
\begin{vmatrix}
 \chi_{1} & \chi_{0}\\
\gamma_{1} \Phi_{1} & {\boxed{\gamma_{0} \Phi_{0}}}
\end{vmatrix}
=\delta_{\chi}^{e}[1]
\end{equation}similarly we can do for the equation (\ref{d4})
\begin{equation}\label{qd2}
 \Phi [1]=
\begin{vmatrix}
 \Phi_{1} & \Phi_{0}\\
\gamma_{1} \chi_{1} & {\boxed{\gamma_{0} \chi_{0}}}
\end{vmatrix}=\delta_{\Phi}^{e}[1]
\end{equation} we have taken $\gamma=\gamma_{0}$, $\chi=\chi_{0}$ and $\Phi=\Phi_{0}$  in order to generalize the transformations in $N$th form.
Further,  we can represent the transformations $\chi [2]$ and $\Phi [2]$ by quasideterminants
 \[\chi [2]=
\begin{vmatrix}
 \chi_{2} & \chi_{1} & \chi_{0}\\
 \gamma_{2} \Phi_{2}& \gamma_{1} \Phi_{1} & \gamma_{0} \Phi_{0}\\
 \gamma^{2}_{2} \chi_{2} & \gamma^{2}_{1} \chi_{1}& {\boxed{\gamma^{2}_{0} \chi_{0}}}
\end{vmatrix}=\delta_{\chi}^{o}[2] \] and
 \[\Phi [2]=
\begin{vmatrix}
 \Phi_{2} & \Phi_{1} & \Phi_{0}\\
 \gamma_{2} \chi_{2} & \gamma_{1} \chi_{1} & \gamma_{0} \chi_{0}\\
 \gamma^{2}_{2} \Phi_{2} & \gamma^{2}_{1} \Phi_{1} & {\boxed{\gamma^{2}_{0} \Phi_{0}}}
\end{vmatrix}=\delta_{\Phi}^{o}[2].\] here superscripts $e$ and $o$  of $\delta$ represent the even and odd order quasideterminants.  The $N$th transformations for $\delta_{\chi}^{o}[N]$ and $\delta_{\Phi}^{o}[N]$ in terms of  quasideterminants are given below
  \[\delta_{\chi}^{o}[N]=
\begin{vmatrix}
 \chi_{N} & \chi_{N-1} & \cdots & \chi_{1} & \chi_{0}\\
\gamma_{N} \Phi_{N} & \gamma_{N-1} \Phi_{N-1} & \cdots & \gamma_{1}\chi_{1}& \gamma_{0} \Phi_{0}\\
 \vdots & \vdots & \cdots & \vdots & \vdots\\
 \gamma^{N-1}_{N} \Phi_{N} & \gamma^{N-1}_{N-1} \Phi_{N-1} & \cdots & \gamma^{N-1}_{1}\chi_{1} & \gamma^{N-1}_{0} \Phi_{0}\\
 \gamma^{N}_{N} \chi_{N} & \gamma^{N}_{N-1} \chi_{N-1} & \cdots & \gamma^{N}_{1} \chi_{1} & {\boxed{\gamma^{N}_{0} \chi_{0}}}
\end{vmatrix}\]and
  \[\delta_{\Phi}^{o}[N]=
\begin{vmatrix}
 \Phi_{N} & \Phi_{N-1} & \cdots& \Phi_{1} & \Phi_{0}\\
 \gamma_{N} \chi_{N} & \gamma_{N-1} \chi_{N-1} & \cdots & \gamma_{1}\chi_{1} & \gamma_{0} \chi_{0}\\
\vdots & \vdots & \cdots & \vdots & \vdots\\
 \gamma^{N-1}_{N} \chi_{N} & \gamma^{N-1}_{N-1} \chi_{N-1} & \cdots & \gamma^{N-1}_{1}\chi_{1} & \gamma^{N-1}_{0} \chi_{0}\\
 \gamma^{N}_{N} \Phi_{N} & \gamma^{N}_{N-1} \Phi_{N-1} & \cdots & \gamma^{N}_{1} \Phi_{1} & {\boxed{\gamma^{N}_{0} \Phi_{0}}}
\end{vmatrix} \]
here $N$ is to be taken as even.
in the same way we can write $N$th quasideterminant representations of $\delta_{\chi}^{e}[N]$ and $\delta_{\Phi}^{e}[N]$.
 \[\delta_{\chi}^{e}[N]=
\begin{vmatrix}
 \chi_{N} & \chi_{N-1} & \cdots & \chi_{1} & \chi_{0}\\
 \gamma_{N} \Phi_{N} & \gamma_{N-1} \Phi_{N-1} & \cdots& \gamma_{1}\chi_{1} & \gamma_{0} \Phi_{0}\\
 \vdots & \vdots& \cdots & \vdots & \vdots\\
 \gamma^{N-1}_{N} \chi_{N} & \gamma^{N-1}_{N-1} \chi_{N-1} & \cdots& \gamma^{N-1}_{1}\chi_{1} & \gamma^{N-1}_{0} \chi_{0}\\
 \gamma^{N}_{N} \Phi_{N} & \gamma^{N}_{N-1} \Phi_{N-1} & \cdots & \gamma^{N}_{1} \Phi_{1} & {\boxed{\gamma^{N}_{0} \Phi_{0}}}
\end{vmatrix}\]and
\[\delta_{\Phi}^{e}[N]=
\begin{vmatrix}
 \Phi_{N} & \Phi_{N-1} & \cdots & \Phi_{1} & \Phi_{0}\\
 \gamma_{N} \chi_{N} & \gamma_{N-1} \chi_{N-1} & \cdots & \gamma_{1}\chi_{1} & \gamma_{0} \chi_{0}\\
 \vdots & \vdots & \cdots & \vdots & \vdots\\
 \gamma^{N-1}_{N} \Phi_{N}& \gamma^{N-1}_{N-1} \Phi_{N-1} & \cdots& \gamma^{N-1}_{1} \Phi_{1} & \gamma^{N-1}_{0} \Phi_{0}\\
 \gamma^{N}_{N} \chi_{N}& \gamma^{N}_{N-1} \chi_{N-1}& \cdots & \gamma^{N}_{1} \chi_{1} & {\boxed{\gamma^{N}_{0} \chi_{0}}}
\end{vmatrix}.\]
Similarly, we can derive the expression for $N$th soliton solution from equation (\ref{d11}) by applying the Darboux transformation iteratively, now consider
\[  v[1]=\Lambda_{1}^{\phi}[1]\Lambda_{1}^{\chi}[1]^{-1} v \Lambda_{1}^{\phi}[1]\Lambda_{1}^{\chi}[1]^{-1}\]
where  \[\Lambda_{1}^{\phi}[1]=\Phi_{1} \]
\[\Lambda_{1}^{\chi}[1] =\chi_{1}\]
this is one fold Darboux transformation. The two fold Darboux transformation is given by
 \begin{equation}\label{it2} 
  v[2]=\phi[1]\chi^{-1}[1]v[1]\phi[1]\chi^{-1}[1].
 \end{equation} 
 We may rewrite the equation (\ref{qd1}) and equation (\ref{qd2}) in the following forms
\[\chi [1]=
\begin{vmatrix}
 \chi_{1} & \chi_{0}\\
\gamma_{1} \Phi_{1} & {\boxed{\gamma_{0} \Phi_{0}}}
\end{vmatrix}
=\Lambda_{2}^{\chi}[2]
\]

 \[\Phi [1]=
\begin{vmatrix}
 \Phi_{1} & \Phi_{0}\\
\gamma_{1} \chi_{1} & {\boxed{\gamma_{0} \chi_{0}}}
\end{vmatrix}=\Lambda_{2}^{\phi}[2].\]
and equation (\ref{it2}) may be written as
\[v[2]=\Lambda_{2}^{\phi}[2] \Lambda_{2}^{\chi}[2]^{-1}\Lambda_{1}^{\phi}[1]\Lambda_{1}^{\chi}[1]^{-1} v \Lambda_{1}^{\phi}[1]\Lambda_{1}^{\chi}[1]^{-1}\Lambda_{2}^{\phi}[2] \Lambda_{2}^{\chi}[2]^{-1}\]
In the same way, we can derive the expression for three fold Darboux transformtion
\[v[3]=\Lambda_{3}^{\phi}[3] \Lambda_{3}^{\chi}[3]^{-1}\Lambda_{2}^{\phi}[2] \Lambda_{2}^{\chi}[2]^{-1}\Lambda_{1}^{\phi}[1]\Lambda_{1}^{\chi}[1]^{-1} v \Lambda_{1}^{\phi}[1]\Lambda_{1}^{\chi}[1]^{-1}\Lambda_{2}^{\phi}[2] \Lambda_{2}^{\chi}[2]^{-1}\Lambda_{3}^{\phi}[3] \Lambda_{3}^{\chi}[3]^{-1}.\]
here
 \[\chi [2]=
\begin{vmatrix}
 \chi_{2} & \chi_{1} & \chi_{0}\\
 \gamma_{2} \Phi_{2}& \gamma_{1} \Phi_{1} & \gamma_{0} \Phi_{0}\\
 \gamma^{2}_{2} \chi_{2} & \gamma^{2}_{1} \chi_{1}& {\boxed{\gamma^{2}_{0} \chi_{0}}}
\end{vmatrix}= \Lambda_{3}^{\chi}[3]\] and
 \[\Phi [2]=
\begin{vmatrix}
 \Phi_{2} & \Phi_{1} & \Phi_{0}\\
 \gamma_{2} \chi_{2} & \gamma_{1} \chi_{1} & \gamma_{0} \chi_{0}\\
 \gamma^{2}_{2} \Phi_{2} & \gamma^{2}_{1} \Phi_{1} & {\boxed{\gamma^{2}_{0} \Phi_{0}}}
\end{vmatrix}=\Lambda_{3}^{\phi}[3].\] 

Finaly, by applying the  transformtion iteratively we can construct the $N$-fold Darboux transformation
\[v[N]=\Lambda_{N}^{\phi}[N] \Lambda_{N}^{\chi}[N]^{-1}\Lambda_{N-1}^{\phi}[N-1] \Lambda_{N-1}^{\chi}[N-1]^{-1}... \Lambda_{2}^{\phi}[2]\Lambda_{2}^{\chi}[2]^{-1} \Lambda_{1}^{\phi}[1]\Lambda_{1}^{\chi}[1]^{-1} v \Lambda_{1}^{\phi}[1]\Lambda_{1}^{\chi}[1]^{-1}\]
   
 \[\Lambda_{2}^{\phi}[2] \Lambda_{2}^{\chi}[2]^{-1}... \Lambda_{N-1}^{\phi}[N-1] \Lambda_{N-1}^{\chi}[N-1]^{-1}\Lambda_{N}^{\phi}[N] \Lambda_{N}^{\chi}[N]^{-1}\]
 by considering the following substitution 
\[\Theta_{N}[N]=\Lambda_{N}^{\phi}[N] \Lambda_{N}^{\chi}[N]^{-1}\]
in above expression, we get
\[v[N]=\Theta_{N}[N]\Theta_{N-1}[N-1]...\Theta_{2}[2]\Theta_{1}[1] v \Theta_{1}[1]\Theta_{2}[2]...\Theta_{N-1}[N-1]\Theta_{N}[N]\]
or
\[v[N]=\Pi^{N-1}_{k=0} \Theta_{N-k}[N-k]V \Pi^{0}_{j=N-1}\Theta_{N-j}[N-j]\]
here we present only the  $N$th expression for odd order quasideterminants  $\Lambda_{N}^{\phi}[N]$ and  $\Lambda_{N}^{\chi}[N]$ 
 \[\Lambda_{2N+1}^{\phi}[2N+1]=
\begin{vmatrix}
 \Phi_{2N} & \Phi_{2N-1} & \cdots& \Phi_{1} & \Phi_{0}\\
 \gamma_{2N} \chi_{2N} & \gamma_{2N-1} \chi_{2N-1} & \cdots & \gamma_{1}\chi_{1} & \gamma_{0} \chi_{0}\\
\vdots & \vdots & \cdots & \vdots & \vdots\\
 \gamma^{2N-1}_{2N} \chi_{2N} & \gamma^{2N-1}_{2N-1} \chi_{2N-1} & \cdots & \gamma^{2N-1}_{1}\chi_{1} & \gamma^{2N-1}_{0} \chi_{0}\\
 \gamma^{2N}_{2N} \Phi_{2N} & \gamma^{2N}_{2N-1} \Phi_{2N-1} & \cdots & \gamma^{2N}_{1} \Phi_{1} & {\boxed{\gamma^{2N}_{0} \Phi_{0}}}
\end{vmatrix} \]
and

 \[\Lambda_{2N+1}^{\chi}[2N+1]=
\begin{vmatrix}
 \chi_{2N} & \chi_{2N-1} & \cdots & \chi_{1} & \chi_{0}\\
\gamma_{2N} \Phi_{2N} & \gamma_{2N-1} \Phi_{2N-1} & \cdots & \gamma_{1}\chi_{1}& \gamma_{0} \Phi_{0}\\
 \vdots & \vdots & \cdots & \vdots & \vdots\\
 \gamma^{2N-1}_{2N} \Phi_{2N} & \gamma^{2N-1}_{2N-1} \Phi_{2N-1} & \cdots & \gamma^{2N-1}_{1}\chi_{1} & \gamma^{2N-1}_{0} \Phi_{0}\\
 \gamma^{2N}_{2N} \chi_{2N} & \gamma^{2N}_{2N-1} \chi_{2N-1} & \cdots & \gamma^{2N}_{1} \chi_{1} & {\boxed{\gamma^{2N}_{0} \chi_{0}}}
\end{vmatrix}\]

\section{Conclusion}
In this paper, I focused on the linear representations of NC P-II equation , I have also constructed its Darboux transformation. Finally I derived its multi-soliton solution in terms of quasideterminants. The further motivations are to explore its other aspects  such that its connection to other integrable equations,  its hierarchy and  Painlev\'e property.
\section{Aknowledgement}
I would like to thank V. Roubtsov, V. Retakh and M. Hassan for their valuable discussions to me during my research work on this paper and especially to M. Cafasso who patiently explained me
the matrix version of Lax pair for NC Painlev\'e II from his paper \cite{BC} with M. Bertola. My special thanks to the university of the Punjab, Pakistan, on funding me for my Ph.D project in France. I am also thankful to S. Meljanac and Z. ˇSkoda for their valuable suggestions and to the franco-croatian cooperation program Egide PHC Cogito 24829NH for a financial support of my visit to Zagreb University.


\noindent
\begin{thebibliography}{99}
\bibitem{r1} P. Painlev\'e, Sur les Equations Differentielles du Second Ordre et d’Ordre Superieur,
dont l’Interable Generale est Uniforme, Acta Math., 25(1902)1-86.
\bibitem{r5}A. N. W. Hone, Painlev\'e test, singularity structure and integrability, arXiv:nlin/0502017v2 [nlin.SI] 22 Oct 2008.
\bibitem{r11}K. Okamoto, in: R. Conte (Ed.), The Painleve Property, One Century Later, CRM Series in Mathematical Physics, Springer, Berlin, 1999, pp. 735-787.
\bibitem{r12}S. P. Balandin, V.V. Sokolov, On the Painlev\'e test for non-abelian equations, Physics letters, A246(1998)267-272.
\bibitem{r4}N. Joshi, The second Painlev\'e hierarchy and the stationarty KdV hierarchy, Publ. RIMS, Kyoto Univ. 40(2004)1039-1061.
\bibitem{MA}N. Joshi, M. Mazzocco, Existence and uniqueness of tri-tronquée solutions of the second Painlevé hierarchy, Nonlinearity 16(2003)427--439 
\bibitem{r13}M. Hamanaka, K. Toda, Towards noncommutative integrable systems, Phys. Lett. A316(2003)77.
\bibitem{r16}L. D. Paniak, Exact noncommutative KP and KdV multi-solitons, hep-th/0105185.
\bibitem{r19} B. A. Kupershmidt, Noncommutative Integrable Systems, in Nonlinear Evolution
Equations and Dynamical Systems, NEEDS 1994, V. Makhankov et al ed-s, , World Scientific 1995, pp. 84-
101.
\bibitem{r22}A. Dimakis, F. M. Hoissen, Noncommutative Korteweg-de Vries equation, Preprint hep-th/0007074, 2000.
\bibitem{r23}M. Legar\'e, Noncommutative generalized NS and super matrix KdV systems from a noncommutative version of (anti-) selfdual Yang-Mills equations, Preprint hep-th/0012077, 2000.
\bibitem{r24}M. Hamanaka, K. Toda , Noncommutative Burgers equation, J. Phys. hep-th/0301213, A36(2003)11981 
\bibitem{e2}A. Dimakis, F. M. Hoissen, With a cole-hopf transformation to solution of noncommutative KP hierarchy in terms of Wronski martices, J. Phys. A40(2007)F32.
\bibitem{1}S. Carillo, C. Schieblod, Noncommutative Korteweg-de Vries and modified Korteweg-de Vries hierarchies via recursion methods, J. Math. Phys.50(2009)073510.
\bibitem{3}I. C. Camero, M. Moriconi, Noncommutative integrable field theories in 2d, Nucl. Phys.B673(2003)437-454
\bibitem{4}A. Dimakis, F. M. Hoissen, The Korteweg-de Vries equation on a noncommutative space-time, Phys. Lett. A278(2000)139-145
\bibitem{5}M. Gurses, A. Karasu, V. V. Sokolov, On construction of of recursion operators from Lax representation, J. Math. Phys.4 0(1985)6473
\bibitem{6}S. P. Balandin, V. V. Sokolov, On the Painleve test for non-Abelian equations, Physics Letters, A246(1998)267-272.
\bibitem{7}V. Retakh, V. Rubtsov, Noncommutative Toda chain , Hankel quasideterminants and Painlev\'e II equation, arXiv:1007.4168v4, math-ph, 2010. 
\bibitem{NCSYM}I. Gelfand, D. Krob, A. Lascoux, B. Leclerc, V. S. Retakh, J.Y. Thibon,  Noncommutative symmetric functions, 	arXiv:hep-th/9407124v1.
\bibitem{Gel Ret} I. Gelfand, S. Gelfand V. Retakh, R. L. Wilson, Quasideterminants , arXiv:math/0208146v4, math.QA, 6 Aug 2004.
\bibitem{r25}J. J. C. Nimmo, collected results on quasideterminants, private notes.
\bibitem{r26}C. R. Gilson,  J. J. C. Nimmo, On a direct approach to quasideterminant solutions of a noncommutative KP equation, J. Phys. A40(2007)no.14,3839--3850
\bibitem{DTI}V. B. Matveev,  M. A.  Salle,  Darboux Transformations and Soliton, Berlin: Springer, 1991. 
\bibitem{DT3} C. Gu, H. Hu,  Z. Zhou, Darboux Transformations in Integrable Systems, Theory and Their Applications
to Geometry , Berlin: Springer ,2005.
\bibitem{DT4}M. Hassan, Darboux transformation of the generalized coupled dispersionless integrable system, J. Phys. A: Math. Theor. 42(2009)065203(11pp)
\bibitem{BC}M. Bertola,  M. Cafasso, Fredholm determinants and pole-free solutions to the noncommutative Painleve' II equation, arXiv:math/1101.3997, math-ph, 20 Jan 2011.

\end{thebibliography}
\end{document}